\pdfoutput=1
\documentclass[twocolumn,superscriptaddress,pra]{revtex4-2}
\usepackage[utf8]{inputenc}
\usepackage[margin=1in]{geometry}
\usepackage{amsmath}
\usepackage{amssymb}
\usepackage{float}
\usepackage{physics}
\usepackage{wasysym}
\usepackage{orcidlink}
 \usepackage{framed}
\definecolor{shadecolor}{named}{lightgray}

\newcommand*{\Radius}{1.0cm}

\usepackage{tikz}
\usetikzlibrary{calc}
\usetikzlibrary{decorations.pathmorphing}
\usetikzlibrary{shapes.misc} 
\usepackage{tikz-cd}
\usepackage{graphicx}
\usepackage{xcolor}

\usepackage{hyperref}

\usepackage{breakcites}

\begin{document}

\title{
Local symmetries and extensive ground-state degeneracy of a 1D supersymmetric fermionic chain
}

\author{Shuyu Zhang}
\email{zhang.s.y.phys@gmail.com}
\affiliation{C. N. Yang Institute for Theoretical Physics, Stony Brook University, USA}
\affiliation{ Department of Physics and Astronomy, Stony Brook University, USA}
\author{Hiroki Sukeno}
\email{hiroki.sukeno@gmail.com}
\affiliation{C. N. Yang Institute for Theoretical Physics, Stony Brook University, USA}
\affiliation{ Department of Physics and Astronomy, Stony Brook University, USA}

\author{Kazuki Ikeda \orcidlink{0000-0003-3821-2669}}
\email{kazuki.ikeda@umb.edu}
\affiliation{Center for Nuclear Theory, Department of Physics and Astronomy, Stony Brook University, Stony Brook, NY, USA}
\affiliation{Department of Physics, University of Massachusetts Boston, Boston, MA, USA}
\author{Tzu-Chieh Wei}
\email{tzu-chieh.wei@stonybrook.edu}
\affiliation{C. N. Yang Institute for Theoretical Physics, Stony Brook University, USA}
\affiliation{ Department of Physics and Astronomy, Stony Brook University, USA}

\begin{abstract}
We study a $1$D supersymmetric (SUSY) hard-core fermion model first proposed by Fendley, Schoutens, and de Boer [Phys. Rev. Lett. {\bf 90}, 120402 (2003)].
We focus on the full Hilbert space instead of a restricted subspace. Exact diagonalization shows the degeneracy of zero-energy states  scales exponentially with size of the system, with a recurrence relation between different system sizes. We solve the degeneracy problem by showing the ground states can be systematically constructed by inserting immobile walls of fermions into the chain. 
Mapping the counting problem to a combinatorial one and obtaining the exact generating function, we prove the recurrence relation on both open and periodic chains. 
We also provide an explicit mapping between ground states, giving a combinatorial explanation of the recurrence relation.
\end{abstract}

\maketitle

\section{Introduction}

Ground state degeneracy (GSD) is a fundamental question in many aspects of physics. In the third law of thermodynamics, it is stated that as the system approaches $T=0$, its entropy approaches a constant (some version states this constant being zero)~\cite{callen1991thermodynamics}. From the perspective of symmetry breaking, it implies that ground states are degenerate, and a change in the degeneracy thus signals phase transition. In topological order, ground-state degeneracy depends on the topology of the underlying manifold~\cite{wen1990topological,wen1990ground}, which can be used to store quantum information with protection~\cite{kitaev2003fault}. 

In many physical models, 
extensive ground-state degeneracy appears, affecting the properties of the phase at low temperatures. 
One of the earliest examples is Pauling's work on the residue entropy of ice~\cite{Pauling1935TheSA}. Others include various frustrated systems, for instance, the frustrated Ising anti-ferromagnetic model on the pyrochlore lattice~\cite{anderson1956ordering} (which led to the discovery of spin ice~\cite{harris1997geometrical}) and $2$D triangular lattice~\cite{PhysRev.79.357}, as well as the random bond Ising model on $2$D square lattice~\cite{JVannimenus_1977} (for 2D systems, see, e.g., the  review~\cite{2004fss..book..229M}). 
Recent studies on topological models and quantum error correction have found fracton models in $3+1$D~\cite{chamon2005quantum,haah2011local,2015PhRvB..92w5136V,vijay2016fracton} that have extensive ground state degeneracies depending on the system size.
Some other relevant quantum systems with extensive degeneracy include supersymmetric (SUSY) model on 2D lattices~\cite{2008EPJB...64..543H}, in which the degeneracy is termed superfrustration, as well as a more recent hidden free fermion model~\cite{2019JPhA...52G5002F}, in which the number of fermion creation/annihilation operators constructed is of order $[2L/3]$, leaving an exponential degeneracy of order $2^{[L/3]}$ in each energy level, where $L$ is the system size.
On a different note, the sharp satisfiability problem (\textbf{$\#$SAT}) is the problem of counting the number of solutions of a given (boolean) satisfiability problem. This is $\#$P-complete, so counting the solutions is, in the worst-case scenario, a hard problem~\cite{doi:10.1137/0208032}. Example problems with exact solutions can be useful as they can benchmark algorithms that aims to solve generic counting problems.

Here, we study the ground-state degeneracy of a particular one-dimensional SUSY fermionic chain by Fendley, Schoutens, and de Boer (FSD)~\cite{fendley2011ground}. 
Various SUSY spin chain models have been studied in the past decades.  For example, given a Majorana SUSY generator (super charge) $Q = Q^\dagger$, with Hamiltonian $\hat{H} = Q^2$, this forms a 0+1D $\mathcal{N}=1$ SUSY algebra~\cite{Witten:1981nf,hori2003mirror,sannomiya2019supersymmetry,2019JPhA...52G5002F,2024ScPP...16..102F}. If the SUSY generator is complex, satisfying $Q^2 = 0$, and the Hamiltonian has the structure 
\begin{align}
    \hat{H} = \{Q^{\dagger},Q\},
\end{align}
then this forms $\mathcal{N}=2$ SUSY algebra~\cite{la2019ground,katsura2020characterization,minar2021supersymmetry,2003JPhA...3612399F, fendley2011ground, 2007JSMTE..02...17F,2018PhRvL.120t6403O,2005PhRvL..95d6403F}. 
It is then easy to see $[Q,\hat{H}] = 0$ due to nilpotency of the SUSY charge. In such models, states with $E>0$ form doublets of the SUSY algebra, while zero-energy states are in the singlets: $Q^\dagger\ket{\psi} = 0$, and there does not exist a state $\ket{\psi'}$ such that $\ket{\psi} = Q\ket{\psi'}$. This says each zero-energy state is an element in the cohomology group $H^f(Q^\dagger)$, where $f$ is the fermion number. It follows that SUSY is spontaneously broken if and only if $H^f(Q^\dagger) = 0$, i.e., a zero dimensional vector space, for all $f$, and the Witten index is given by $W = \sum_{f}(-1)^f {\rm dim} H^f(Q^\dagger)$~\cite{WITTEN1982253}.

The FSD SUSY lattice model on a periodic chain~\cite{PhysRevLett.90.120402} has the following SUSY generator
\begin{equation}
Q^\dagger = \sum_{i} P_{i-1}c_i^\dagger P_{i+1} , 
\end{equation}
where $c_i,c^\dagger_i$ are fermionic annihilation and creation operators (at site $i$)  satisfying  $\{c_i,c^\dagger_j\}=\delta_{ij}$. $P_i$ is a projection onto the un-occupied state $P_i = (1-n_i)$ with $n_i = c_i^\dagger c_i$. Each term $P_{i-1}c_i^\dagger P_{i+1}$ creates a hard-core fermion since the neighbors of $i$ must be empty in order to create a fermion on site $i$. The full Hamiltonian is
\begin{align} \label{eq:hamiltonian}
  \hat{H}  =\sum_{i=1}^N[P_{i-1}(c^\dagger_ic_{
i+1}+c^\dagger_{i+1}c_{i})P_{i+2}+P_{i-1}P_{i+1}],
\end{align}
where the periodic boundary condition is assumed.
The Hilbert space on which $Q^\dagger$ acts is a subspace of the full Hilbert space, which has the restriction that no neighboring sites can be both occupied. 
This restriction is well-defined as $Q^\dagger$ maps this restricted Hilbert space to itself. This model exhibits many interesting properties. It was shown by FSD~\cite{fendley2011ground} that, for a chain with $N=3p$, there are two ground states (in the restricted Hilbert space) with their Witten index being $W =
2(-1)^p$, and they carry a translation eigenvalue  $t=(-1)^N e^{\pm i \pi/3}$. For $N = 3p \pm1$, there is a unique ground state with  $W=(-1)^p$ and $t=(-1)^{N-1}$. All ground states have a fermion number $f =\text{int}
((N+1)/3)$.

\begin{table}[t]
    \begin{tabular}{|l|l|}
    \hline
    \begin{tabular}{l}boundary \\ condition 
    \end{tabular}& degeneracy sequence from $L=4$ \\
    \hline
     periodic & 
    \begin{tabular}{l}10,12,24,44,66,114,202,332,560,964,\\
    1626,2746,...
    \end{tabular}\\
     \hline
    open (0,0) & 2,4,4,8,16,24,40,72,120,...\\
    open (0,1) &  2,4,8,12,20,36,60,100,172,...\\
   open (1,1) & 4,6,10,18,30,50,86,146,246,... \\
   \hline
    \end{tabular}
    \caption{Ground-state degeneracy for the supersymmetric fermionic chain in the 
 unrestricted Hilbert space with different boundary conditions, starting with $L=4$. For open boundary conditions, it can be easily checked that they satisfy the recurrence relation: $a_n=a_{n-1}+2 a_{n-3}$. This recurrence relation holds for the periodic boundary condition when $n\ne 3m-1$ (where $m$ is an integer), otherwise $a_{3m-1}=a_{3m-2}+2a_{3(m-1)-1}-2$. The ground-state degeneracy in this table was obtained by exact diagonalization of the corresponding spin chain after the Jordan-Wigner transformation; see Appendix~\ref{app:spin}.}\label{tb:degeneracy}
    \end{table}

Our work here considers the ground-state degeneracy of the same fermionic chain in the unrestricted Hilbert space, where we do not forbid nearest-neighbor occupancy. When this constraint is relaxed, we expect the ground-state degeneracy to increase. Such curiosity led us to compute the degeneracy using exact diagonalization for small systems, for both periodic and open boundary conditions (see Appendix~\ref{app:spin} for the corresponding spin Hamiltonian used in the diagonalization), with the results presented in Table~\ref{tb:degeneracy}, where there are some empirical recurrence relations emerging from the degeneracy among different system sizes. 
A similar recurrence relation was observed for the Nicolai model in Ref.~\cite{2017PhRvD..95f5001S} and later proved in
Ref.~\cite{la2019ground} using tools from homological algebra. 
In this paper, we aim to derive the corresponding recurrence relations for the FSD model, as well as providing an explicit mapping, hence obtaining the exact ground-state degeneracy.
Technically, we have found that the framework of the spectral sequence, exploited in the original work of Ref.~\cite{PhysRevLett.90.120402}, was very useful in understanding the ground-state structure in the restricted Hilbert space. These serve as building blocks of the ground states in the unrestricted space. Once this is understood, we can catapult the result to the unrestricted case by labeling ground states with certain integer partitions and, hence, turning it into a simple combinatorial problem. 
As will be shown, the partitions label immobile walls in different ground states, which are products of occupied states $\ket{1\cdots 1}$, and there is no entanglement across the walls.
Furthermore, we have found that the open boundary conditions were easier to deal with, and their results paved the way for the ground-state degeneracy in the periodic case without the Hilbert space restriction. We summarize our results as follows,
\begin{shaded}
\noindent {\bf Result:} Ground state degeneracy of the Hamiltonian~\eqref{eq:hamiltonian} on the open chain follows the recursion relation 
        $a_n=a_{n-1}+2 a_{n-3}$, while on the periodic chain it follows the relation
        $a_n=a_{n-1}+3a_{n-3}-a_{n-4}-2a_{n-6}$.
           
\end{shaded}

The remainder of this paper is organized as follows.
In Sec.~\ref{sec:SpetralSquence}, we present a minimal version of the spectral sequence, which is a key tool for us. In Sec.~\ref{sec:OBC}, we first tackle the open boundary conditions, which have three inequivalent cases. Solving these cases helps to build our path towards the periodic boundary condition, which we discuss in Sec.~\ref{sec:PBC}. We make some concluding remarks in Sec.~\ref{sec:Conclusion}.

\section{Spectral Sequence}\label{sec:SpetralSquence}
Readers who are familiar with the spectral sequence for bi-graded complexes can skip this section. The key outcome is Eq.~\eqref{eq:hnk}. For completeness, we present a simplified version of spectral sequence following Section 14 of \cite{Bott1982DifferentialFI}, which is very useful in studying the cohomology of $Q^\dagger$ in the restricted Hilbert space.
Given a double complex $C = \bigoplus_{s,t \in \mathbb{N}}C^{s,t}$,
\[
\begin{tikzcd}
&\cdots &\cdots \\
\cdots \arrow[r,"\delta"]& C^{s,t+1} \arrow[u,"d"]\arrow[r, "\delta^{s,t+1}"] 
& C^{s+1,t+1} \arrow[u,"d"]\arrow[r, "\delta"]
& \cdots\\
\cdots \arrow[r,"\delta"]
&C^{s,t} \arrow[u,"d^{s,t}"] \arrow[r, "\delta^{s,t}"]
& C^{s+1,t} \arrow[u,"d^{s+1,t}"]\arrow[r, "\delta"]
& \cdots\\
&\cdots \arrow[u,"d"] 
&\cdots \arrow[u,"d"] 
\end{tikzcd}
\]
where $C^{s,t}$ are vector spaces. For our applications, we adopt the convention that the diagram anticommutes, meaning $\delta^{s,t+1}d^{s,t}+d^{s+1,t}\delta^{s,t}=0$, because $\delta$ and $d$ will be fermionic operators. The labels $s$ and $t$ will be fermion numbers on two properly chosen sublattices, and $C^{s,t}$ are each fermion number sectors. We may assume the double complex is bounded in both $s$ and $t$. We can define an associated chain complex $D^n: C^n\rightarrow C^{n+1}$ with
\begin{align}
    &C^n = \bigoplus_{s+t=n}C^{s,t},\\
    &D^n = \bigoplus_{s+t=n} (\delta^{s,t}+ d^{s,t}).
\end{align}
One can check $D^{n+1}D^n = 0$. Denoting the total complex to be $K = \bigoplus_n C^n$, we wish to compute the cohomology of this complex $H^n(K)$.
The idea is to take a finite sequence of subspaces $K=K_0\supset K_1 \supset \cdots \supset K_N\supset 0$ (for our application $N$ could be the size of the lattice). 
This gives a similar sequence for cohomology groups $H^n(K)=H^n(K_0)\supset H^n(K_1) \supset \cdots \supset H^n(K_N)\supset 0$. Using this sequence $H^n(K)$ can be approximated step by step, and eventually given exactly by a sum of quotients
\begin{align}
    H^n(K) = \bigoplus_{p=0}^N H^n(K_p)/H^n(K_{p+1}).
\end{align}
We note for vector spaces the above holds, while for modules over rings it is not necessarily true. We illustrate this explicitly. 
Let us choose
\begin{align}
    K_p = \bigoplus_{s\geq p;t}C^{s,t}
    \hspace{25pt}
    p\geq 0,
\end{align}
and define $K_p = K$, $\forall p<0$ and denote the inclusion map as $i:K_p\rightarrow K_{p-1}$. Note each $K_p$ is itself a chain complex with coboundary operator $D$, and $i$ induces an inclusion on cohomology groups $i^*: H^n(K_p) \rightarrow H^n(K_{p-1})$. For each $p$ we have a short exact sequence (SES) of chain complexes
\begin{align}
    0\rightarrow K_p \xrightarrow{i_p} K_{p-1} \xrightarrow{j_p} K_{p-1}/K_p \rightarrow 0,
\end{align}
where $j$ is the quotient map. Such an SES gives rise to a long exact sequence (LES)
\begin{align} 
    \cdots \rightarrow H^n(K_p) &\xrightarrow{i_p^*} H^n(K_{p-1}) \xrightarrow{j_{p}^*} H^n(K_{p-1}/K_p) \nonumber \\
    & \xrightarrow{\partial_p} H^{n+1}(K_p) \rightarrow \cdots \label{eq:les}
\end{align}
Note it is not hard to see $K_{p-1}/K_{p}$ is just the chain complex $C^{p-1,*}$ with coboundary $d$ (the vertical map). This shows
\begin{align}
    H^{n}(K_{p-1}/K_p) = H_d^{p-1,n-p+1}(K),
\end{align} the cohomology at coordinate $(p-1,n-p+1)$ with respect to $d$. The boundary map $\partial_p$ is just $D = \delta+ d$, and we have omitted the superscripts to simplify notation. 
However, $d$ is $0$ acting on $H_d(K)$, and therefore $\partial_p$ is simply $\delta^{p-1,n-p+1}$ (the horizontal map). 
We denote $A_1^{p,n}\equiv H^n(K_p)$ and $B_1^{p-1,n-p+1}\equiv H^n(K_{p-1}/K_p)= H_d^{p-1,n-p+1}(K)$.  

From Eq.~(\ref{eq:les}) we can write a LES involving only $B_1$: 
\begin{align}
    \cdots\rightarrow B_1^{p-2,n-p+1} &\xrightarrow{j_{p}^* \circ \partial_{p-1}} B_1^{p-1,n-p+1} \nonumber\\
    &\xrightarrow{j_{p+1}^* \circ \partial_p} B_1^{p,n-p+1} \rightarrow \cdots
\end{align}
The exactness is due to $\partial\cdot j^* = 0$. Define
\begin{align}
    &B_2^{*,*} = H(B_1^{*,*}),\\
    &A_2^{p,n} = i^* (A_1^{p,n})\subset A_1^{p-1,n}.
\end{align}
Now we recall that $\partial$ is simply the map $\delta:H_d^{s,t}(K) \rightarrow H^{s+t+1}(K_{s+1})$, and $j^*:H^{s+t+1}(K_{s+1})\rightarrow H_d^{s+1,t}(K)$ is the quotient map, so we see $j^*\circ \partial ([a]) = [\delta(a)]$, $\forall [a] \in H_d^{s,t}(K)$, and this gives
\begin{align}
    B_2^{s,t} = H^{s,t}_\delta (H_d(K)),
\end{align}
which is the cohomology on $H_d(K)$ with respect to $\delta$. 

Now we can derive another LES from Eq.~(\ref{eq:les}) involing $A_2$ and $B_2$ 
\begin{align}
    \cdots \rightarrow A_2^{p,n} \xrightarrow{i^*} A_2^{p-1,n} &\rightarrow B_2^{p-1,n-p+1} \nonumber\\
    & \xrightarrow{\partial} A_2^{p+1,n+1} \rightarrow \cdots
\end{align}
The specific form of $\partial$ is not important here. One can repeat this process indefinitely. However in our case, after $N+1$ steps $A_{N+1}^{p,n}\subset A_1^{p-N,n}=H^{n}(K)$, $\forall p,n$, and the map $i^*: A_{N+1}^{p,n} \rightarrow A_{N+1}^{p-1,n}$ is exactly the inclusion map between the two subsets of $H^{n}(K)$. Since $i^*$ is injective, exactness implies $\partial=0$, and the LES breaks into many SES
\begin{align}
     0\rightarrow A_{N+1}^{p,n} \rightarrow A_{N+1}^{p-1,n} \rightarrow B_{N+1}^{p-1,n-p+1}\rightarrow 0,
\end{align}
from which it follows $B_{N+1}^{p-1,n-p+1} = A_{N+1}^{p-1,n}/A_{N+1}^{p,n}$. Since we have a sequence of inclusions $H^n(K)\supset A_{N+1}^{0,n}\supset \cdots \supset A_{N+1}^{N,n}$, we have shown
    $H^n(K) = \bigoplus_{s+t=n} B_{N+1}^{s,t}$.
In our application below, we will see in fact there is only one non-zero $B_2$, and hence the sequence stabilizes after $B_2$, so we have $B_2^{s,t} = B_{N+1}^{s,t}$, $\forall s,t$. This gives
\begin{align}
    \label{eq:hnk}
    H^n(K) = \bigoplus_{s+t=n}H^{s,t}_\delta (H_d(K)).
\end{align}

\section{Open boundary conditions}
\label{sec:OBC}
\subsection{Restricted Hilbert Space}

Given the 1D lattice with size $L$, we label the sites with $\{0,1,\cdots \Tilde{L},\Tilde{L}+1\}$ and set $L = \Tilde{L}+2$. The SUSY generator we are interested in is 
\begin{align}
    Q = \sum_{i=1}^{\Tilde{L}} P_{i-1}c_i P_{i+1}.
\end{align}
Let us focus on the restricted Hilbert space.
The two operators $n_0$ and $n_{\Tilde{L}+1}$ commute with $Q$, and hence there are 4 sectors $(n_0,n_{\Tilde{L}+1})$ labeling different boundary conditions: $n_{0,\Tilde{L}+1} = 0,1$. The essential ingredient in both restricted and unrestricted cases is the $(0,0)$ sector. That is, sites $0$ and $\Tilde{L}+1$ are unoccupied. The ground state degeneracy is computed via the rank of the cohomology group of $Q^\dagger$ in the restricted Hilbert space. 
Spectral sequence is especially useful in this case. 

We give an explicit example using $\Tilde{L} = 3n$. The first step is to choose sublattices: we take sublattice 2 (SL2) to be $\{2,5,\cdots, 3n-1\}$, and the rest to be sublattice 1 (SL1). Use $K^{f_1,f_2}$ to label the restricted Hilbert space with $f_1$ and $f_2$ fermions on SL1 and SL2 respectively. 
The differential operators are
\begin{align}
    &\delta=\sum_{i\in SL1}P_{i-1}c_i^\dagger P_{i+1} , \\
    &d=\sum_{i\in SL2}P_{i-1}c_i^\dagger P_{i+1} .
\end{align}
By prescription, we first compute the cohomology of $d$, which gives $B_1$. These are states such that for each site in SL2, at least one of its neighbors is occupied. 
There are two non-zero terms $B_1^{n,0} = {\rm span}\{\ket{i}\,|\, i= 0,\cdots, n\}$ and $B_1^{n+1,0} = {\rm span}\{|\hat{i}\rangle\,|\, i = 1,\cdots, n\}$, where
\begin{align*}
    &\ket{i}=\begin{tikzpicture}[baseline={([yshift=-.5ex]current bounding box.center)},
    >=Stealth,
  shorten >=1pt,
  auto,
  node distance=0.3cm,
  thick,
  main node/.style={draw,shape=circle,fill,inner sep=1pt}]
    \node [main node][label = 0] (0) at (-0.3,0) {};
    \node [red] (1) at (0,0) {$\times$};
    \node [main node] (2) at (0.3,0) {};
    \node [main node] (3) at (0.6,0) {};
    \node [red] (4) at (0.9,0) {$\times$};
    \node [main node] (5) at (1.2,0) {};
    \node (6) at (1.8,0) {$\cdots$};
    \node [red] (7) at (2.4,0) {$\times$};
    \node [main node][label = $3i-1$] (8) at (2.8,0) {};
    \node [main node] (9) at (3.1,0) {};
    \node [main node] (10) at (3.4,0) {};
    \node [main node] (11) at (3.7,0) {};
    \node [red] (12) at (4.0,0) {$\times$};
    \node (13) at (4.6,0) {$\cdots$};
    \node [main node] (14) at (5.2,0) {};
    \node [red] (15) at (5.5,0) {$\times$};
    \node [main node] (16) at (5.8,0) {};
    \node [main node] (17) at (6.1,0) {};
    \node [red] (18) at (6.4,0) {$\times$};
    \node [main node] [label=$L-1$] (L) at (6.7,0) {};
    \draw
    (0) -- (2) -- (3) --(5) -- (6) -- (8) -- (9) --(10) --(11) -- (13) -- (14)--(16)--(17)-- (6.8,0) ;
    \end{tikzpicture}\\
   &|\hat{i}\rangle=\begin{tikzpicture}[baseline={([yshift=-.5ex]current bounding box.center)},
    >=Stealth,
  shorten >=1pt,
  auto,
  node distance=0.3cm,
  thick,
  main node/.style={draw,shape=circle,fill,inner sep=1pt}]
    \node [main node][label = 0] (0) at (-0.3,0) {};
    \node [red] (1) at (0,0) {$\times$};
    \node [main node] (2) at (0.3,0) {};
    \node [main node] (3) at (0.6,0) {};
    \node [red] (4) at (0.9,0) {$\times$};
    \node [main node] (5) at (1.2,0) {};
    \node (6) at (1.8,0) {$\cdots$};
    \node [red] (7) at (2.4,0) {$\times$};
    \node [main node][label = $3i-1$] (8) at (2.8,0) {};
    \node [red] (9) at (3.1,0) {$\times$};
    \node [main node] (10) at (3.4,0) {};
    \node [main node] (11) at (3.7,0) {};
    \node [red] (12) at (4.0,0) {$\times$};
    \node (13) at (4.6,0) {$\cdots$};
    \node [main node] (14) at (5.2,0) {};
    \node [red] (15) at (5.5,0) {$\times$};
    \node [main node] (16) at (5.8,0) {};
    \node [main node] (17) at (6.1,0) {};
    \node [red] (18) at (6.4,0) {$\times$};
    \node [main node] [label=$L-1$] (L) at (6.7,0) {};
    \draw
    (0) -- (2) -- (3) --(5) -- (6) -- (8) --(10) --(11) -- (13) -- (14)--(16)--(17)-- (6.8,0) ;
    \end{tikzpicture}
\end{align*}
in which a red cross \textcolor{red}{$\times$} indicates occupation. Next we compute the cohomology of the horizontal boundary operator $\delta:B_1^{n,0}\rightarrow B_1^{n+1,0}$, which is associated to the following chain complex
\begin{align} 
    \cdots \rightarrow 0 &\rightarrow B_1^{n,0} \xrightarrow{\delta} B_1^{n+1,0} \rightarrow 0 \rightarrow \cdots \label{eq:les}.
\end{align} 
The action is
\begin{align}
    &\ket{i} \mapsto (-1)^i(|\hat{i}\rangle+|\hat{i}+1\rangle),\\
    &\ket{0} \mapsto \ket{\hat{1}},
    \hspace{35pt}
    \ket{n} \mapsto (-1)^n\ket{\hat{n}}.
\end{align}
It is straightforward to see ${\rm Ker}\delta = \ket{0} + \ket{1} + \cdots + \ket{n}$, and ${\rm Im}\delta = B_1^{n+1,0}$. Hence 
the only nonzero terms is $B_2^{n,0} = {\rm span}\{\ket{0} + \cdots +\ket{n}\}$. As discussed in the last section, the spectral sequence stabilizes here, and we find
one zero-energy state with $n$ fermions. 

This simple exercise can be carried out for $\Tilde{L}=3n+1$ and $\Tilde{L}=3n+2$, and we state the results here. For $\Tilde{L}=3n+1$, we find $B_1^{n,0} = {\rm span}\{\ket{i}\,|\, i = 0,\cdots, n\}$, and $B_1^{n+1,0} = {\rm span}\{|\hat{i}\rangle\,|\, i = 1, \cdots, n+1\}$, where the diagram for $|i\rangle$ and $|\hat{i}\rangle$, $\hat{i}\leq n$, is similar to the previous one, except with one additional unoccupied appended at the end. 
The diagram for the additional $|\hat{n}+1\rangle$ is as follows
\begin{align}
    |\hat{n}+1\rangle=\begin{tikzpicture}[baseline={([yshift=-.5ex]current bounding box.center)},
    >=Stealth,
  shorten >=1pt,
  auto,
  node distance=0.3cm,
  thick,
  main node/.style={draw,shape=circle,fill,inner sep=1pt}]
    \node [main node][label = 0] (0) at (-0.3,0) {};
    \node [red] (1) at (0,0) {$\times$};
    \node [main node] (2) at (0.3,0) {};
    \node [main node] (3) at (0.6,0) {};
    \node [red] (4) at (0.9,0) {$\times$};
    \node [main node] (5) at (1.2,0) {};
    \node (6) at (1.8,0) {$\cdots$};
    \node [red] (7) at (2.4,0) {$\times$};
    \node [main node][label = $3n-1$] (8) at (2.8,0) {};
    \node [main node] (9) at (3.1,0) {};
    \node [red] (10) at (3.4,0) {$\times$};
    \node [main node] (11) at (3.7,0) {};
    \draw
    (0) -- (2) -- (3) --(5) -- (6) -- (9) --(3.8,0)  ;
    \end{tikzpicture}
\end{align}
The boundary operator $\delta:B_1^{n,0}\rightarrow B_1^{n+1,0}$ acts as
\begin{align}
     \ket{i} \mapsto (-1)^i(|\hat{i}\rangle+|\hat{i}+1\rangle), 
    \hspace{25pt}
    \ket{0} \mapsto \ket{\hat{1}}.
\end{align}
There are no zero-energy states. For $\Tilde{L}=3n+2$, $B_1^{n,0} = {\rm span}\{\ket{i}\,|\, i = 0,\cdots, n\}$ and $B_1^{n+1,0} = {\rm span}\{|\hat{i}\rangle\,|\, i = 0, \cdots,  n+1\}$. 
Here we append one extra unoccupied site at the beginning, and the vector $|\hat{0}\rangle$ is
\begin{align}
    |\hat{0}\rangle=\begin{tikzpicture}[baseline={([yshift=-.5ex]current bounding box.center)},
    >=Stealth,
  shorten >=1pt,
  auto,
  node distance=0.3cm,
  thick,
  main node/.style={draw,shape=circle,fill,inner sep=1pt}]
    \node [main node][label = 0] (0) at (-0.3,0) {};
    \node [red] (1) at (0,0) {$\times$};
    \node [main node] (2) at (0.3,0) {};
    \node [main node] (3) at (0.6,0) {};
    \node [red] (4) at (0.9,0) {$\times$};
    \node [main node] (5) at (1.2,0) {};
    \node (6) at (1.8,0) {$\cdots$};
    \node [red] (7) at (2.5,0) {$\times$};
    \node [main node] (8) at (2.8,0) {};
    \node [main node][label = $3n$] (9) at (3.1,0) {};
    \node [red] (10) at (3.4,0) {$\times$};
    \node [main node] (11) at (3.7,0) {};
    \node [main node] (12) at (4.0,0) {};
    \draw
    (0) -- (2) -- (3) --(5) -- (6) -- (11) --(4.1,0)  ;
    \end{tikzpicture}
\end{align}
The boundary operator $\delta:B_1^{n,0}\rightarrow B_1^{n+1,0}$ acts as
\begin{align}
     \ket{i} \mapsto (-1)^i(|\hat{i}\rangle+|\hat{i}+1\rangle).
\end{align}
There is one ground state with $f=n+1$.

The above cohomology calculations allow us to count the number of zero-energy state of the Hamiltonian. However, we do not know a concise way to explicitly write down the ground-state wavefunctions for all $L$. For small sizes we can easily  construct them. For instance, for $L=4$: 
\begin{align}
    \ket{\psi}_4=\ket{0100}-\ket{0010}
\end{align}
and for $L=5$ 
\begin{align}
    \ket{\psi}_5=\ket{01000}+\ket{00010}-2\ket{00100}.
\end{align} 

We also include some special cases: for $L=3$, there is no ground states; for $L=2$ and $L=1$, each has one ground state with $f=0$, which are just $\ket{\psi}_2=\ket{00}$ and $\ket{\psi}_1=\ket{0}$, respectively. 

To summarize, we find one zero-energy state for $L\equiv 1$ (mod 3) with fermion number $[L/3]$; one zero-energy state for $L\equiv 2$ (mod 3) with fermion number $[L/3]$; and no zero-energy states for $L\equiv 0$ (mod 3). In particular, when $L$ is a multiple of 3, SUSY is spontaneously broken.

Finally, other results for other sectors follow from the result for $(0,0)$ sector. The $(1,0)$ and $(0,1)$ (parity reflection related) with size $L$ is obtained from $(0,0)$ sector with size $L-1$ by appending a $\ket{1}$ to an end, and $(1,1)$ sector with size $L$ is obtained from $(0,0)$ sector with size $L-2$ by appending $\ket{1}$ to both ends.
Hence in $(1,0)$ or $(0,1)$ sector, SUSY is broken on $L\equiv 1$ (mod $3$), and in $(1,1)$ sector, SUSY is broken on $L\equiv 2$ (mod $3$).

\subsection{Unrestricted Hilbert Space: (0,0) Sector}

Next we move to the unrestricted Hilbert space. In this section we first focus on the $(0,0)$ sector. Using this as an example, we explain our methodology, which can easily be applied not only to other other sectors in open boundary condition, but also to periodic chain.

Exact diagonalization shows the GSD grows exponentially with system size, satisfying a recurrence relation. For instance in $(0,0)$ sector we find $4,4,8,16,24,40,72,120, \cdots$ starting at $L=5$. We show this integer sequence can be packed into a generating function which is simply a rational function.

Essential ingredient is to realize the symmetry $n_in_{i+1}\sim CZ$. This says if there are 2 fermions stuck together ($n_in_{i+1}=1$), then they stay like this during dynamics. In particular, we can split the Hilbert spaces into eigen-spaces of these projectors, where each eigen-space is characterized by a number of immobile ``walls," which are consecutive occupied sites $\ket{\cdots 111 \cdots}$. These walls cut off all interactions between its two sides: fermions cannot tunnel through the walls, and there are no entanglement across the walls. On the other hand, a segment in between 2 walls must have its 2 boundaries unoccupied, and must not contain any walls within itself. Hence it is exactly a zero-energy state in the restricted Hilbert space, and belongs to the $(0,0)$ sector. Looking at the $(0,0)$ sector first, a generic ground state looks like the following:
\begin{align}
    &\begin{tikzpicture}
        \draw (0,0) -- (0.2,0);
        \draw [line width=0.5mm, red ]  (0.2,0) -- (1,0);
        \draw (1,0) -- (1.4,0);
        \draw [line width=0.5mm, red ]  (1.4,0) -- (1.7,0);
        \draw (1.7,0) -- (2.4,0);
        \draw [line width=0.5mm, red ]  (2.4,0) -- (2.9,0);
        \draw (2.9,0) -- (3.3,0);
        \draw [line width=0.5mm, red ]  (3.3,0) -- (3.8,0);
        \draw (3.8,0) -- (4.0,0);
    \end{tikzpicture}\nonumber\\
    &\ket{\psi} = \ket{-} \otimes \ket{\textcolor{red}{-}}\otimes \cdots \otimes \ket{\textcolor{red}{-}} \otimes \ket{-}
\end{align}
where $\ket{\textcolor{red}{-}}$ is are the walls consists a number of occupied sites $\ket{1\cdots 1}$, and $\ket{-}$ is the unique ground state in the restricted Hilbert space in $(0,0)$ sector, which we studied in previous section.

This structure gives us a systematic way to write down all zero-energy states. In particular, we are able to count all such states in a systematic way. This is done by rephrasing into the following combinatorial problem: given $L\in \mathbb{Z}$ (the size of the chain), we find pairs of partitions $(\lambda,\mu)$, $\lambda = \{\lambda_i\}$, $\mu = \{\mu_j\}$ and $|\lambda|+|\mu| = L$ ($\lambda$ label the sizes of the open chains, and $\mu$ label the thicknesses of the walls). They are subjected to the following requirements:
\begin{enumerate}
    \item Length of partitions satisfies $l(\lambda) = l(\mu) +1$
    \item $\lambda_i \not\equiv 0$ (mod 3)
    \item $\mu_i \geq 2$
\end{enumerate}
Each pair contributes $d(\lambda,\mu)$:
\begin{align}
    d(\lambda,\mu) = |S(\lambda)|\cdot |S(\mu)| .
\end{align}
where $S(\lambda)$ is the set of all distinct permutations of $\lambda$, and $|\cdot|$ is the size of the set. For example, take two partitions of $3 $, $\lambda=\{2,1\}$, $\mu = \{3\}$, then $|S(\lambda)|=2$ and $|S(\mu)| = 1$, so $d(\lambda,\mu) = 2$. The ground state degeneracy is simply a sum from all allowed pairs of partitions. Denoting $P$ to be the set of all such allowed partitions, and $GSD(L,00)$ to be the GSD on size $L$ chain in $(0,0)$ sector, then
\begin{align}
    GSD(L;00) = \sum_{(\lambda,\mu)\in P} d(\lambda,\mu).
\end{align}
We have formulated the GSD problem into a problem of counting partitions.

The generating function is a useful tool for such problems. Given an integer sequence $\{a_n\}_{n=0}^\infty$, we can pack this sequence into a polynomial $a(x) = \sum_{n=0}^{\infty} a_nx^n$. For instance, recall the generating function for all Young diagrams is \cite{s11}
\begin{align}
    y(x) = \prod_{n=1}^{\infty}(1+x^n+x^{2n} + \cdots) = \prod_{n=1}^{\infty}\frac{1}{1-x^n}.
\end{align}
In our case, we also need to take into account of distinct permutations of a partition. For instance, given a partition $\lambda = \{3,3,3,3,2,2,1\}$, the number of distinct permutations are $7!/(4!2!1!)$. The numerator is the factorial of $l(\lambda)$, and the denominator is a product of factorials of the multiplicity of $3$, $2$, and $1$ respectively. To this end, we introduce an additional perimeter $a$ to record the length of the partition, and first define
\begin{align*}
    \Tilde{Y}(x,a) = \prod_{n=1}^\infty (1+ ax^n + \frac{1}{2!}a^2x^{2n} + \cdots) = \prod_{n=1}^\infty e^{ax^n}.
\end{align*}
The generating function $\Tilde{y}(x)$ can be expressed in terms of $\Tilde{Y}(x,a)$ by
\begin{align}
    \Tilde{y}(x) = \left.\left(\sum_{k=0}^\infty \frac{\partial^k}{\partial a^k} \Tilde{Y}(x,a)\right)\right|_{a=0}.
\end{align}
We can apply this directly to our problem.

In analogy of $\Tilde{y}(x)$, the generating function $g_{00}(x)$ counting GSD in $(0,0)$ sector can be written in the following way:
\begin{align}
    &G(x,a,b)
    =\prod_{\substack{m=1 \\ m\not\equiv 0 \, (mod\, 3)}}^{\infty}e^{ax^m}\prod_{n=2}^{\infty} e^{bx^n},\\
    & g_{00}(x) = \left.\left(\sum_{i=0}^\infty\frac{\partial^{i+1}}{\partial a^{i+1}}
    \frac{\partial^i}{\partial b^i}
    G(x,a,b)\right)\right|_{a,b = 0}.
\end{align}
where $a$ and $b$ records the length of $\lambda$ and $\mu$ respectively. There is always one more derivative in $a$ because $l(\lambda)$ is always larger than $l(\mu)$ by $1$.
The GSD in the $(0,0)$ sector is nicely encoded in the coefficients of the power series:
\begin{align}
    GSD(L;00) = \left.\frac{1}{L!}\partial^L_x g(x)\right|_{x=0}.
\end{align}
Next we show $g_{00}(x)$ can be resummed, using
\begin{align}
    &\partial_a G(x,a,b) = \frac{x+x^2}{1-x^3} G(x,a,b) , \\
    &\partial_b G(x,a,b) = \frac{x^2}{1-x} G(x,a,b) .
\end{align}
We find
\begin{align}
    g_{00}(x) &= \sum_{i=0}^\infty \left(\frac{x+x^2}{1-x^3}\right)^{i+1}\left(\frac{x^2}{1-x}\right)^i\\
    =&  -\frac{x(x^2-1)}{(1-x^3)(1-x)}\sum_{i=0}^{\infty}\left(\frac{x^3(x+1)}{(1-x^3)(1-x)}\right)^i\\
    =& -\frac{x(x^2-1)}{(1-x^3)(1-x)}\frac{1}{1-\frac{x^3(x+1)}{(1-x^3)(1-x)}}\\
    =&\frac{x(1-x^2)}{1-x-2x^3} .
\end{align}
We show a few low order terms of the expansion of $g(x)$:
\begin{align}
    g_{00}(x) = x + x^2 + 2 x^4 + 4 x^5 + 4 x^6 + \mathcal{O}(x^7) .
\end{align}
The denominator of this rational function $2x^2 + x -1$ represents a recurrence relation
\begin{align*}
    GSD(L;00) = GSD(L-1;00) + 2\cdot GSD(L-3;00) .
\end{align*}
This is a relation between GSD in different system sizes. 

It is not hard to give it a combinatorial interpretation. Denoting $K(L,00)$ to be the set of all ground states in $(0,0)$ sector, we define the following explicit map from the disjoint union $K(L-1,00)\bigsqcup K(L-3,00)\bigsqcup K(L-3,00)$ to $K(L,00)$
\begin{enumerate}
    \item given a ground state in $K(L-1,00)$, insert a fermion (tensor with $\ket{1}$) into the leftmost wall, obtaining a state in $K(L,00)$.
    \item given a ground state in $K(L-3,00)$, insert a $\ket{110}$ on the left of the leftmost wall, obtaining a state in $K(L,00)$.
    \item given a ground state in $K(L-3,00)$, if the leftmost wall has thickness $>2$, then modify the leftmost wall $\ket{111\cdots 1}$ into $\ket{110011\cdots 1}$, i.e. inserting $\ket{100}$ on the right of the first fermion, obtaining a state in $K(L,00)$.
    \item given a ground state in $K(L-3,00)$, if the leftmost wall has thickness 2, then enlarging the open chain between leftmost and second leftmost wall by size 3, obtaining a state in $K(L,00)$.
\end{enumerate}
One can check this is indeed a bijection between the two sets: pick $\ket{\psi}\in K(L,00)$, first bullet point above enumerates the cases where the thickness of the leftmost wall in $\ket{\psi}$ is at least $3$. Second and third point enumerates the cases where the leftmost wall has thickness $2$, and the size of the open chain between the leftmost and second leftmost wall is $1$ and $2$, respectively. Finally, the 4th point accounts for all the rest. In this way, the pre-image of $\ket{\psi}$ exists and is unique.

\subsection{Other open boundary conditions}
\noindent {\bf (1,0) or (0,1) sector}. These $2$ cases are parity reflections of each other, and the counting is exactly the same, and therefore we focus on $(1,0)$ sector without loss of generality.
Here, the leftmost site is always occupied. Hence there is always a wall starting on the leftmost site. A generic ground state looks like
\begin{align}
    &\begin{tikzpicture}
        \draw [line width=0.5mm, red ]  (-0.5,0) -- (-0.2,0);
        \draw (-0.2,0) -- (0.2,0);
        \draw [line width=0.5mm, red ]  (0.2,0) -- (1,0);
        \draw (1,0) -- (1.4,0);
        \draw [line width=0.5mm, red ]  (1.4,0) -- (1.7,0);
        \draw (1.7,0) -- (2.4,0);
        \draw [line width=0.5mm, red ]  (2.4,0) -- (2.9,0);
        \draw (2.9,0) -- (3.3,0);
        \draw [line width=0.5mm, red ]  (3.3,0) -- (3.8,0);
        \draw (3.8,0) -- (4.0,0);
    \end{tikzpicture} \nonumber \\
    &\ket{\psi} = \ket{\textcolor{red}{-}}\otimes \ket{-} \otimes \cdots \otimes \ket{\textcolor{red}{-}} \otimes \ket{-}
\end{align}
However there is one difference: the first wall starting on the leftmost site is allowed to have thickness $1$. We hereby separate the calculation into 2 cases. First, the first wall has size $1$. Such states are obtained from states in $K(L-1,00)$ by simply adding an extra fermion on the left boundary (tensor with $\ket{1}$). Its contribution to generating function $g_{10}(x)$ is simply $xg_{00}(x)$.
Second, the first wall has thickness $2$ or above. The analysis is identical to the $(0,0)$ case. We look for appropriate pairs of partitions $(\lambda,\mu)$, except the first condition is modified to $l(\lambda) = l(\mu)$, i.e. the two partitions now have the same size. Corresponding contribution to $g_{10}(x)$ is
\begin{align} 
\label{eq:g10_secondpart}\left.\left(\sum_{i=1}^\infty\frac{\partial^{i}}{\partial^{i} a}
    \frac{\partial^i}{\partial^i b}
    G(x,a,b)\right)\right|_{a,b = 0} = \frac{x^2}{1-x} g_{00}(x) .
\end{align}
In total, we find
\begin{align}
    g_{01}(x)=g_{10}(x) &= xg_{00}(x) + \frac{x^2}{1-x}g_{00}(x)\\
    &= \frac{x^2(1+x)}{1-x-2x^3} .
\end{align}
The denominator indicates the GSD has the same recurrence relation as in the $(0,0)$ sector, and a similar map can be defined.

\smallskip
\noindent{\bf (1,1) sector}. In this case both ends must be occupied, and both the leftmost and rightmost wall are allowed to have thickness $1$. We separate the calculation into three cases. First, both first and last wall has thickness $1$: the contribution to $g_{11}(x)$ is simply $x^2g_{00}(x)$. Second, only one of the walls has thickness $1$: this contributes $2xg_{10}(x)$.
where the factor of $2$ takes into account that either leftmost or rightmost wall can have thickness $1$. The third possibility is that both walls have thickness $2$ or above. The contribution to $g_{11}(x)$ is
\begin{align}
\label{eq:tildeg11}
 \left.\left(\sum_{i=0}^\infty\frac{\partial^{i}}{\partial^{i} a}
    \frac{\partial^{i+1}}{\partial^{i+1} b}
    G(x,a,b)\right)\right|_{a,b = 0} = \frac{1-x^3}{x+x^2}g_{10}(x) .
\end{align}
In total, we find
\begin{align}
    g_{11}(x) &= x^2g_{00}(x) + 2x g_{10}(x) + \frac{1-x^3}{x+x^2}g_{10}(x) \\
    &=  \frac{x^2(1+x+2x^2)}{1-x-2x^3} .
\end{align}
Again, the GSD satisfies the same recurrence relation.

\section{Periodic boundary condition}
\label{sec:PBC}

It turns out that we can obtain most of the ground states on the periodic chain in the unrestricted Hilbert space by suitably joining the cases considered above in open boundary conditions. The remaining ground states not obtained this way are from the Bethe ansatz solutions of Fendley, Schoutens, and de Boer.

We aim to find the generating function $g_p(x)$ periodic boundary condition: 
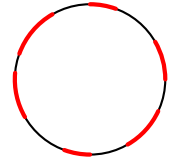
\begin{align*}
&\begin{tikzpicture}[ultra thick, line cap=round,line join=round]
\coordinate (Origin) at (0,0);
\coordinate (1) at (0:\Radius);
\coordinate (2) at (70:\Radius);
\coordinate (3) at (120:\Radius);
\coordinate (4) at (175:\Radius);
\coordinate (5) at (250:\Radius);
\coordinate (6) at (300:\Radius);
\draw [black, thick] (Origin) circle[radius=\Radius];
\draw [red] (Origin) ++ (1) arc[radius=\Radius,start angle=0,end angle=30];
\draw [red] (Origin) ++ (2) arc[radius=\Radius,start angle=70,end angle=90];
\draw [red] (Origin) ++ (3) arc[radius=\Radius,start angle=120,end angle=160];
\draw [red] (Origin) ++ (4) arc[radius=\Radius,start angle=175,end angle=210];
\draw [red] (Origin) ++ (5) arc[radius=\Radius,start angle=250,end angle=270];
\draw [red] (Origin) ++ (6) arc[radius=\Radius,start angle=300,end angle=335];
\end{tikzpicture}\\
\ket{\psi} =  \ket{\textcolor{red}{-}}&\otimes\ket{-} \otimes \cdots \otimes \ket{\textcolor{red}{-}} \otimes \ket{-}
\end{align*}
Our strategy is the same as before. We cut the circle open, and count partitions. Conversely, we may construct ground states on periodic chain by gluing open chains. 

To glue states from different open-boundary conditions, we begin with the simplest case of (1,1). We can simply join the two ends to form the nearest neighbors. The contribution to $g_p(x)$ from this part is $g_{11}(x)$.

Next, we consider the (0,1)/(1,0) sector. In this case, only the configuration where the rightmost/leftmost wall has thickness at least $2$ can be joined to obtain a ground state configuration on the open chain, simply because the walls in the periodic case cannot have thickness $1$. The contributes to $g_p(x)$ is therefore twice the expression in Eq.~(\ref{eq:g10_secondpart}), giving rise to $\frac{2x^2}{1-x^2}g_{00}(x)$.

The contributions from the (0,0) condition are slightly more complicated due to the mod $3$ constraint open chain segments. Specifically, when the segments on two ends are joined, the total length must not be a multiple of $3$. 
These states can be built from those ground states in the third contribution to the $(1,1)$ sector in Eq.~(\ref{eq:tildeg11}) by attaching segments on both ends. Define this as the bulk generating function
\begin{align}
    g_p^{\rm bulk}(x)=\frac{1-x^3}{x+x^2}g_{10}(x).
\end{align}
Denoting the lengths of the segments on two ends by $l_1$ (left end) and $l_2$ (right end) respectively, this leads to three cases:
\begin{enumerate}
    \item $l_1\equiv l_2\equiv 1 ({\rm mod}\,3)$. The contribution to $g_p(x)$ is
    \begin{align*}
    \left(\sum_{n=0}^\infty x^{3n+1}\right)^2 g_p^{\rm bulk}(x) = \frac{x^2}{(1-x^3)^2} g_p^{\rm bulk}(x) .
    \end{align*}
    \item $l_1\equiv l_2\equiv 2 ({\rm mod}\,3)$. The contribution to $g_p(x)$ is
    \begin{align*}
    \left(\sum_{n=0}^\infty x^{3n+2}\right)^2 g_p^{\rm bulk}(x) = \frac{x^4}{(1-x^3)^2} g_p^{\rm bulk}(x) .
    \end{align*}
    \item either $l_1\not\equiv 0 ({\rm mod}\,3)$ and $l_2\equiv 0 ({\rm mod}\,3)$, or vice versa. The contribution to $g_p(x)$ is
    \begin{align*}
        2\left(\sum_{\substack{m=1 \\ m\not\equiv 0 \, (mod\, 3)}}^{\infty}x^m \right) \left(\sum_{n=1}^\infty x^{3n}\right)g_p^{\rm bulk}(x) 
    \end{align*}
    which resums to 
    \begin{align}
        2\frac{x^4+x^5}{(1-x^3)^2}g_p^{\rm bulk}(x) .
    \end{align}
\end{enumerate}
These count all ground state configurations on a periodic chain with at least one wall. We also need to account for the ground states with no walls, which are the solutions in \cite{PhysRevLett.90.120402}. 
For $L\equiv 0$ (mod $3$) there are two Bethe ansatz solutions, and for $L\not\equiv 0$ (mod $3$) there are two solutions. Their contribution to $g_p(x)$ is
\begin{align}
    2\sum_{n=0}^\infty x^{3n} + \sum_{\substack{m=1 \\ m\not\equiv 0 \, (mod\, 3)}}^{\infty}x^m = \frac{2+x+x^2}{(1-x^3)} .
\end{align}
Summing all the contributions, the final expression for the generating function in the periodic case is thus
\begin{equation}
  g_p(x)=  -x-\frac{-1+x+x^2}{-1+x^3}+\frac{-3+2x}{-1+x+2x^3}.
\end{equation}
Similar to the previous obc case, we see the recurrence is (due to $(-1 + x^3) (-1 + x + 2 x^3)=1-x-3x^3+x^4+x^6$ )\begin{equation}
    a_n=a_{n-1}+3a_{n-3}-a_{n-4}-2a_{n-6}.
\end{equation}
Rewriting the above equation gives
\begin{equation}
    a_n-a_{n-1}-2a_{n-3}=a_{n-3}-a_{n-4}-2a_{n-6},
\end{equation}
which is a slight variant of the open boundary case.

\section{Conclusion}
\label{sec:Conclusion}
We have calculated ground-state degeneracy for a 1D supersymmetric fermionic model with both open and periodic boundary conditions without restricting the Hilbert space. We did this by reformulating a problem on integer partitions.  
It turns out that the degeneracy in all cases satisfies similar recurrence relations, and thus, their exact values can be easily obtained. We also give a combinatorial interpretation of this recurrence relation by defining an explicit mapping.
A previous work on the periodic case with a restricted Hilbert space found that the Witten index is nonzero for all system sizes. We found that for the open boundary conditions with a restricted Hilbert space, there is no zero-energy ground state for $L=3n$. In the full Hilbert space, there are many zero-energy states, the number of which can be obtained from our solutions.

The essential ingredient is the local $n_in_{i+1}$ symmetry, which equals to $CZ_{i,i+1}$ after Jordan-Wigner transformation up to a constant, although the former is non-invertible, while the latter is unitary. This breaks down the problem into studying system on an open chain with ``restricted space", and here restricted means $n_{i}n_{i+1} = 0$ for all $i$. This can be applied to analyze any system with local $n_{i}n_{i+1}$ symmetry. However if the Hamiltonian is not positive definite, one might need to track the energies on each segment.
We remark that, for the Nicolai model~\cite{nicolai1976supersymmetry}, existence of local symmetries was pointed out and they were used to characterize its extensive ground states in Ref.~\cite{katsura2020characterization,moriya2018ergodicity}. It would be interesting to understand the relation between their symmetries in the Nicolai model and ours in the FSD model.

It is also proved in \cite{la2019ground} that the recurrence relation can in fact be refined into involving distinct fermion number sectors. We also observed this from periodic case with small system sizes. In fact our explicit mapping can serve as a proof of this recurrence relation on the open chain. It would be interesting to see if this can be deduced from the perspective of partition combinatorics using generating function as well.

Moreover, since the local $n_in_{i+1}$ is a symmetry of the Hamiltonian, the entire spectrum can be arranged in terms of the wall configurations. In particular, it would be interesting to know whether or not there is a finite gap separating the degenerate ground state from the lowest excitations and how one can place this model in the classification of phases. 
Intriguingly, the immobile walls are similar to blockades in systems which conserves dipole moment, and it would be also interesting to study  Hilbert space fragmentation~\cite{sala2020ergodicity,khemani2020localization,2022PhRvX..12a1050M} in the FSD model; see Ref.~\cite{Surace2020weakergodicity} for a study in this direction.

\begin{acknowledgments}
    We thank Paul Fendley for useful discussions, particularly for pointing out that the walls are immobile. This work was supported by the National Science Foundation under Award No. PHY 2310614.
\end{acknowledgments}

\appendix
\section{Spin Hamiltonian}\label{app:spin}
We can convert the fermionic Hamiltonian into the spin representation using the Jordan-Wigner transformation~\cite{Jordan:1928wi}, 
\begin{align}
\begin{aligned}
n_i&=\frac{1+Z_i}{2}\\
c^\dagger_n&=\frac{X_n+iY_n}{2}\prod_{j=1}^{n-1}Z_j\\
c_n&=\frac{X_n-iY_n}{2}\prod_{j=1}^{n-1}Z_j
\end{aligned}
\end{align}
where $Z_i$ is the local Pauli $Z$ at site $i$. They obey the anti-commutation relation $\{c_i,c^\dagger_j\}=\delta_{ij}$. Using the anti-commutation relations $\{X,Z\}=\{Y,Z\}=0$, we find 
\begin{align}
\begin{aligned}
c^\dagger_ic_{i+1}+c^\dagger_{i+1}c_{i}=-\frac{1}{2}(X_iX_{i+1}+Y_iY_{i+1}).
\end{aligned}
\end{align}
Therefore, the spin representation of the Hamiltonian for the open boundary condition is (without imposing any condition on the two sides at the boundary)
is 
\begin{align}
\begin{aligned}
\label{eq:Pauli}
    \hat{H}=&-\frac{1}{8}\sum_{i=1}^{\Tilde{L}}(1-Z_{i-1})(X_iX_{i+1}+Y_iY_{i+1})(1-Z_{i+2})\\
    &+\frac{1}{4}\sum_{i=1}^N(1-Z_{i-1})(1-Z_{i+1}).
\end{aligned} 
\end{align}
The specific open  $(n_{0},n_{\Tilde{L}+1})$ 
boundary condition can be imposed by adding to the Hamiltonian two additional terms $(I+(-1)^{n_{0}}Z_0) + (I+(-1)^{n_{\Tilde{L}+1}}Z_{\Tilde{L}+1})$. 
To ensure the ground states obtained in the spin representation correspond to those in the restricted Hilbert space, we also need to add terms to penalize nearest-neighbor occupation: To obtain the ground states for restricted Hilbert space, we need to add penalty terms in enforce that no two neighbors are simultaneously occupied: $n_{i-1} n_i=(1+Z_{i-1})(1+Z_{i})/4$.

For the periodic boundary condition, the term associated with the one $(X_L X_1 + Y_L Y_1)$ needs to be multiplied by a factor $-\eta=- \prod_{i=1}^L Z_j$, which is the negative of the fermion or spin-even-odd parity.

The above spin Hamiltonians allow us to diagonalize numerically to obtain the ground-state degeneracy for small system sizes.

\raggedright
\bibliography{ref}
\end{document}